\documentclass[printnumbers,amsmath,amssymb,twocolumn]{revtex4-2}
\usepackage{graphicx}
\usepackage{color}
\usepackage{float}
\usepackage{epstopdf}
\usepackage{hyperref}
\usepackage{booktabs}

\begin{document}

\title{On  the ratio between longitudinal and transverse Ioffe-Regel frequencies in  glasses}  

 \author{ Licun Fu$^{1}$,  Mingyu Zhu$^{1}$,  Xinyu Chang$^{1}$,   and Lijin Wang$^{1,*}$ \\
$^{1}${School of Physics, State Key Laboratory of Opto-Electronic Information Acquisition and Protection Technology, Anhui University, Hefei 230601, China}\\
$^{*}$ Corresponding author,  Email: lijin.wang@ahu.edu.cn\\
}

\begin{abstract}

The temperature dependence of thermal conductivity in glasses differs characteristically from that in crystals, and is largely controlled by how   sound  waves  are  damped. The Ioffe–Regel (IR) frequency  sets the high-frequency limit of well-defined  sound waves. Although the transverse IR frequency has been established to coincide with the boson peak frequency and reported to be far below the longitudinal IR frequency, the quantitative relation between longitudinal and transverse IR frequencies has remained unknown. Here we examine this relation  in   two- (2D) and three-dimensional (3D) model glasses with vastly different stability.    We observe  that,  in 3D glasses,  the  longitudinal-to-transverse IR frequency ratio is approximately equal to the ratio of the  low-frequency transverse to longitudinal sound attenuation coefficient, whereas this correspondence is not observed in  2D glasses.   Moreover,  we find that the glass stability strongly controls the longitudinal-to-transverse IR frequency  ratio  in both 2D and 3D glasses:  the ratio decreases significantly with increasing glass stability and approaches unity for the most stable glasses under study.  Accordingly, in sufficiently stable glasses, the longitudinal sound  attenuation should be comparable to the transverse attenuation, and therefore cannot be treated as negligible,   challenging  common assumptions adopted especially in  theoretical studies.   Our work is  another demonstration of the  glass stability as a key parameter for investigating glass properties, and cautions against simply extrapolating observations  from poorly annealed  glasses to  stable  glasses.


\end{abstract}

\date{\today}

\maketitle

\section{Introduction}

When quenched sufficiently rapidly, nearly all liquids can be supercooled and ultimately form glasses~\cite{Berthier_2011_RMP}. One well-known distinction between glasses and their crystalline counterparts of the same chemical substances is the anomalous yet universal temperature dependence of the thermal conductivity $\kappa(T)$~\cite{Ramos_book,experim_Zeller1971,thoery_Anderson1972,review_thermal2002,Phillips_JLTP_1972}.  At  temperatures below approximately 2K, glasses  exhibit $\kappa \sim T^2$, in contrast to the $\kappa \sim T^3$  scaling observed  in crystals. This universal anomaly in glasses is  usually rationalized within the framework of two-level tunneling model~\cite{thoery_Anderson1972,Phillips_JLTP_1972}.
In the intermediate-temperature regime of roughly 2–20K, the thermal conductivity of glasses is nearly temperature independent, which is  markedly different from its behavior in crystals. To account for this plateau in $\kappa(T)$, it is essential to reveal both the spectrum of low-frequency excess vibrational modes~\cite{Wang-ropp,lerner-JCP-review} and the  mechanism underlying sound attenuation~\cite{Szamel-SA,Wang2019SMattenuation,Szamel-SA,lerner2019JCP,Wang2020softmater,Fupre,Szamel-PRE-2025,Ikeda_pre2018,Lemaitre-NM,Baggioli-JPCM-2022,Szamel-jcp-2022,Szamel-jcp-2025}.  Intriguingly, accumulating  evidence has also suggested a correlation between the  sound attenuation and excess vibrational modes~\cite{Schrimacher_prl2007,bp_shintani_NM2008,Wang2019SMattenuation,Massimo_prl2021}.  

In crystalline solids, the low-frequency vibrational density of states $D(\omega)$ obeys the Debye law  $D(\omega)=A_{\rm D}\omega^{d-1}$, where $A_{\rm D}$ is the Debye level~\cite{Kittel}, and $d$ is the spatial dimension. In  glasses, however, the low-frequency vibrational density of states  typically exceeds the Debye prediction~\cite{Wang-ropp,lerner-JCP-review,mizuno_bigsystem_dos_pnas,wlj_dos_nc2019,Wang2021prl,Xu2010EPL,lerner_prl2016,wang_3d_jcp2022,wang_2d_jcp2023,Wang-CPB-4D-dos,lerner_dos_234d_prl2018,patrick-PRL,Wang_prl2014,Massimo_prl2021,XuDing}.
This excess contribution manifests as a distinct peak in the plot of $D(\omega)/\omega^{d-1}$ against $\omega$,  known as the boson peak, occurring at a characteristic frequency commonly referred to as the boson peak frequency~\cite{Nakayama-ROPP-2002,Schrimacher_prl2007,Schirmacher-SciRep}.   It is noteworthy  that the statistical properties of excess vibrational modes have attracted substantial research attention over the past decade~\cite{Wang-ropp,lerner-JCP-review}.

The boson peak frequency was reported to coincide approximately with the transverse Ioffe–Regel (IR) frequency $\omega_{\rm T,IR}$, in two- (2D) and three-dimensional  (3D) glasses, which  implies  that the excess vibrational modes associated with the boson peak are predominantly transverse in character~\cite{bp_shintani_NM2008,BPorigin_yuanchao-NP}.  At the IR frequency, the phonon mean free path becomes comparable to  the phonon wavelength, and hence phonons cease to be well-defined  at frequencies above the IR frequency~\cite{IR}. With respect to the longitudinal IR frequency $\omega_{\rm L,IR}$, it is commonly reported that $\omega_{\rm L,IR} > \omega_{\rm T,IR}$ in most glasses~\cite{Ikeda_pre2018, bp_shintani_NM2008,  Monaco-pnas-2009,BPorigin_yuanchao-NP,xipeng-PRL}. A notable exception is observed in network silica glasses, where simulations~\cite{silica-model-1,silica-model-2} confirm the equality  $\omega_{\rm L,IR} = \omega_{\rm T,IR}$. This equality   may be due to the strong directional covalent bonding in silica,  or   silica glass-forming liquids being classified as “strong” in the sense of the Angell plot~\cite{Staley-JCP-2015,Berthier_2011_RMP,Angell_1991_JNCS,Angell_1995_Science}.  One pertinent question is whether the ratio \(\beta_{\rm L/T,IR} = \omega_{\rm L,IR} / \omega_{\rm T,IR}\) can be systematically tuned in simple model glasses that do not incorporate additional complexities such as the covalent bonding. To our knowledge, there has been  one simulation study by Nie \textit{et al.}~\cite{Nie-Frontier} attempting to address this issue in the lattice model.  Nie \textit{et al.} introduced specific types of disorder into an otherwise perfect crystal,  and demonstrated that $\beta_{\rm L/T,IR}$ can indeed be modified in this manner.  However,  it remains unclear whether and  how $\beta_{\rm L/T,IR}$ can be tuned  in realistic glasses where multiple types of  disorder coexist.

Moreover,  although the sound attenuation in glasses has attracted considerable  interest~\cite{Szamel-jcp-2025}, most existing theoretical and numerical studies have mainly focused on the transverse  branch~\cite{Ikeda_pre2018,Vogel_2023_PRX}.  This  may primarily arise from the numerical observation in poorly annealed glasses that the longitudinal sound attenuation  is small—often negligible—compared with its transverse counterpart~\cite{Ikeda_pre2018, bp_shintani_NM2008,  Monaco-pnas-2009,BPorigin_yuanchao-NP,xipeng-PRL}. For instance, 
 within the framework of the generalized Debye theory~\cite{Ikeda_pre2018}, only transverse sound attenuation is  taken into account, whereas the longitudinal attenuation is neglected under the assumption that it is significantly smaller than $\Gamma_{\rm T}(\omega)$.   
 
Additionally, recent simulation studies~\cite{Monaco-pnas-2009, Wang2019SMattenuation,Wang2020softmater,Fupre} have  suggested  that the ratio of the low-frequency transverse sound  attenuation coefficient to the low-frequency longitudinal one obeys $\Gamma_{\rm T}(\omega)/\Gamma_{\rm L}(\omega)=\beta_{\rm  T/L,\Gamma}$,  where $\beta_{\rm  T/L,\Gamma}$ is a system-specific constant.  To our knowledge, in  all relevant simulation studies,  $\beta_{\rm  T/L,\Gamma}$ is determined empirically so that the $\Gamma_{\rm T}(\omega)$ data can be collapsed onto the $\Gamma_{\rm L}(\omega)$ data by rescaling.  However, it remains unknown whether $\beta_{\rm  T/L,\Gamma}$ can be determined in a more physical  manner.   Alternatively,  one open question is what control parameter governs $\beta_{\rm  T/L,\Gamma}$.

 In this work, our numerical simulations of model glass formers suggest that $\beta_{\rm L/T,IR}$ systematically decreases as glass stability increases in both 2D and 3D glasses. For poorly annealed 3D glasses away  from the jamming transition, $\beta_{\rm L/T,IR}$ is approximately insensitive to variations in number density and to the choice of interaction potential model. For poorly annealed 2D glasses away  from the jamming transition, $\beta_{\rm L/T,IR}$ likewise exhibits negligible dependence on density, but seems to be sensitive to the interaction potential model. Intriguingly,  we observe $\beta_{\rm L/T,\Gamma} \approx \beta_{\rm L/T,IR}$ in 3D glasses, whereas this correspondence is not consistently satisfied in 2D glasses. These results imply that the effective control parameter governing both $\beta_{\rm L/T,\Gamma}$ and $\beta_{\rm L/T,IR}$ is the same in 3D glasses, but differs in 2D glasses.  Additionally,  we demonstrate that  the fragility of glass-forming liquids has nearly no correlation with the value of  $\beta_{\rm L/T,IR}$ in corresponding glasses.

\section{Simulation details}

 We simulated  model glass formers where particles interact via the  inverse power law (IPL)  potential~\cite{Ninarello_2017_PRX}.   Specifically,  the interaction between  particles $i$ and $j$  is   defined as
\[
U(r_{ij})=\Bigg[\left(\frac{\sigma_{ij}}{r_{ij}}\right)^{n}+c_{0}+c_{2}\left(\frac{r_{ij}}{\sigma_{ij}}\right)^{2}+c_{4}\left(\frac{r_{ij}}{\sigma_{ij}}\right)^{4}\Bigg]\,G(r_{ij}^{c}-r_{ij}),
\]
where \(G(r)\) denotes the Heaviside step function, \(r_{ij}\) is the  separation between particles $i$ and $j$, and the  interaction cutoff distance \(r_{ij}^{c}=1.25\,\sigma_{ij}\).  Particle diameters are continuously distributed according to a distribution of \(P(\sigma) \sim \frac{1}{\sigma^{3}}\) over the interval \textbf{\(0.73 \leq \sigma \leq 1.62\)}. The cross-diameter is defined by a non-additive mixing rule, i.e.,  \(\sigma_{ij} = \frac{\sigma_i + \sigma_j}{2}\left(1 - 0.2\,\lvert \sigma_i - \sigma_j \rvert\right)\). All particles are assumed to have identical mass.    The additional polynomial terms are introduced to enforce smoothness of \(U(r_{ij})\) up to its second derivative at  \(r_{ij}^{c}\). We consider two exponents:  \(n=6\) (IPL6), and  \(n=12\) (IPL12).  

Additionally, we extracted  literature data for glass formers with the harmonic potential  (denoted as HARM)~\cite{Ikeda_pre2018}, and the Lenard-Jonnes potential (LJ)~\cite{Monaco-pnas-2009}.  

Zero-temperature ($T=0$) glassy configurations were generated by instantaneously quenching finite-temperature equilibrium  configurations to $T=0$ through energy minimization using the fast inertial relaxation engine algorithm~\cite{fire}. 
 The number density is defined as $\rho=N/L^d$, where $L$ is the side length,  and $N$ is the total number of particles. $N$ in the IPL glasses ranges from $20000$ to $1000000$, depending on $\rho$,  $T_{ \rm p}$, and $d$.   Here,  $T_{ \rm p}$ is the  parent temperature from which the $T=0$ glasses are quenched. 
Independent equilibrated configurations  at each low $T_{ \rm p}$ were prepared using swap Monte Carlo method~\cite{Ninarello_2026_JCP,Ninarello_2017_PRX}. 
 For  the (3D, IPL12, $\rho=1$) glasses~\cite{Ninarello_2017_PRX},  $T_{\rm p}$  changes from  the onset temperature ($T_o \approx 0.2$) of slow dynamics  to $0.062$  which is about 86\% of the estimated glass transition temperature ($T_{\rm g} \approx 0.072$).  For  the (2D, IPL12, $\rho=1$) glasses~\cite{Berthier_2019_NC},  $T_{\rm p}$  changes from $0.4$ which  is above the onset temperature ($T_o \approx 0.25$) of slow dynamics  to $0.03$  which is about 36\% of the estimated glass transition temperature ($T_{\rm g} \approx 0.082$). 
 For  poorly annealed glasses quenched from  high-temperature liquids or randomly packed configurations,  their parent temperatures  are referred  to as  $T_{\rm p}=\infty$ in this study. Specifically,  for literature HARM data~\cite{Ikeda_pre2018},  glasses were generated by quenching quickly    randomly packed configurations to  $T=0$, and hence the parent temperature is referred to as $T_{\rm p}=\infty$. For literature LJ data~\cite{Monaco-pnas-2009}, glasses were obtained by quenching high-temperature liquids to extremely low temperatures at a high cooling rate, and we  denote as well the parent temperature as $T_{\rm p}=\infty$ in this system.

\begin{figure}[t]
\includegraphics[width=0.45\textwidth]{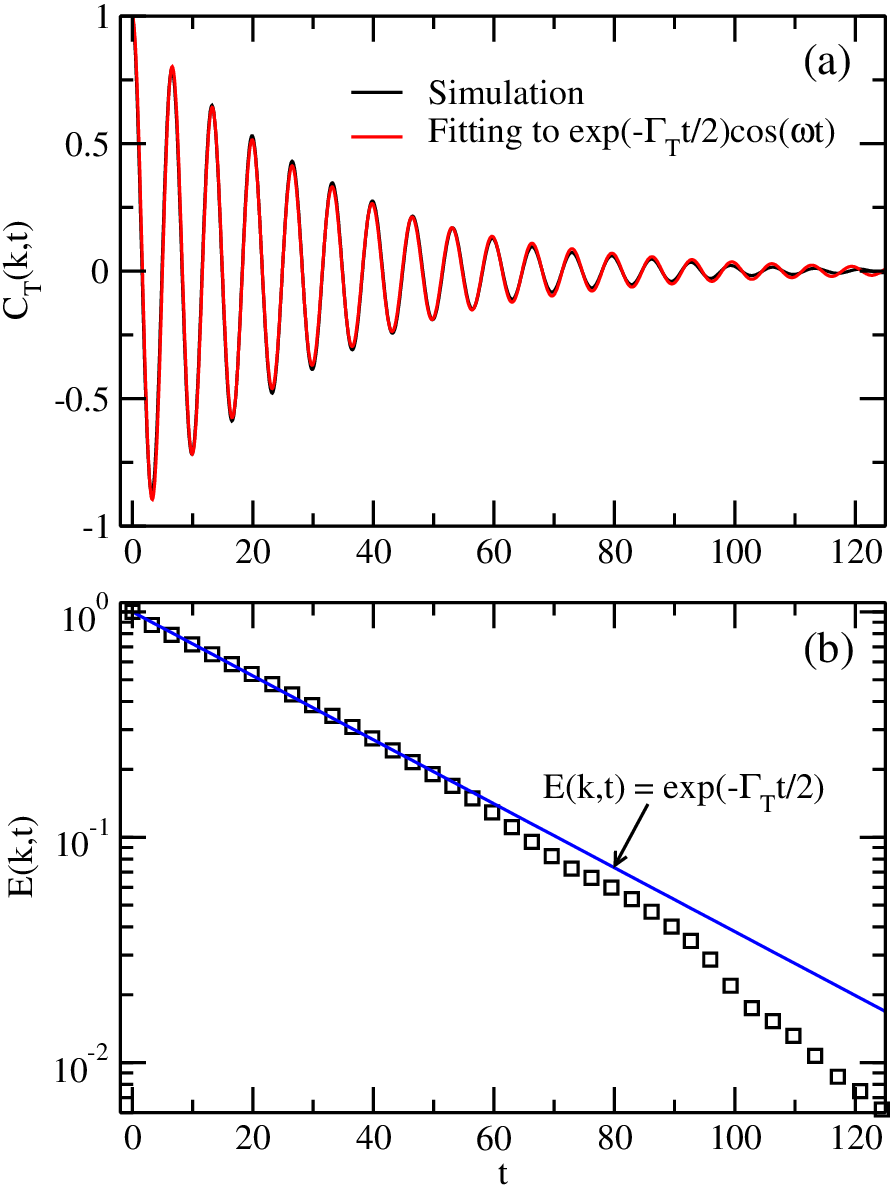}
\caption{\label{fig:envelope} (a) Velocity auto-correlation function $C_{\rm T}(k,t)$ corresponding to a transverse excitation at  a small wavevector $k=0.216$ in the  (3D, IPL12, $\rho=1.5$, $T_{\rm p}=\infty$) glasses.  (b) The temporal evolution of the envelope $E(k,t)$  corresponding to the maxima of the absolute value $|C_{\rm T}(k,t)|$.  The blue line indicates a fit to $E (k,t) =\exp(-\Gamma_{\rm T} t/2)$.
}
\end{figure}

\begin{figure}[h]
\includegraphics[width=0.45\textwidth]{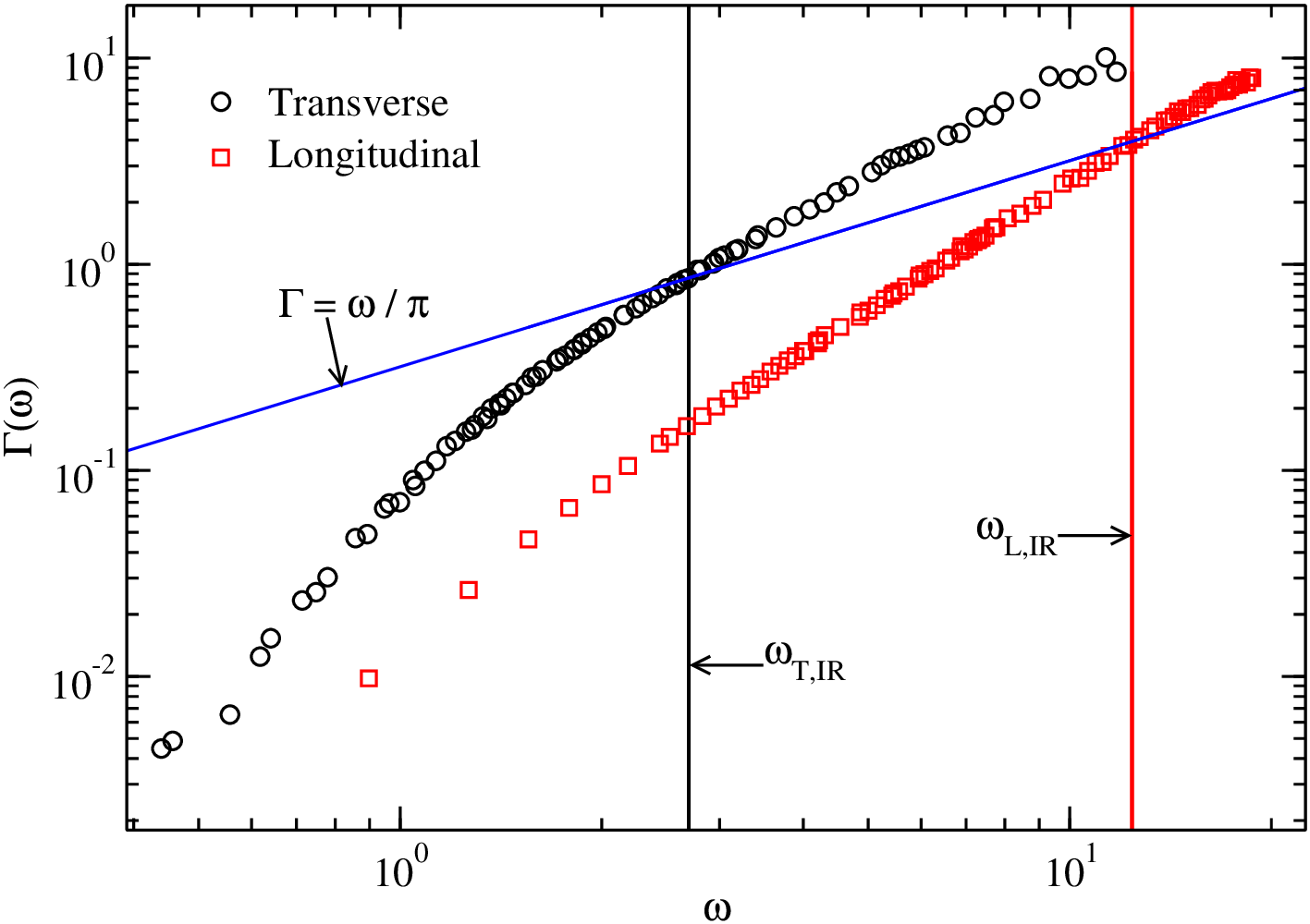}
\caption{\label{fig:OmegaIR} Illustration of the procedure used to determine IR frequencies in the (3D, IPL12, $\rho=1.5$, $T_{\rm p}=\infty$) glasses. The transverse IR frequency $\omega_{\rm T,IR}$ is defined by the intersection of the curve $\Gamma_{\rm T}(\omega)$ with the reference line $\Gamma=\omega/\pi$. An analogous construction is used to determine the longitudinal IR frequency $\omega_{\rm L,IR}$. 
}
\end{figure}

 \begin{figure*}[t]
\includegraphics[width=0.9\textwidth]{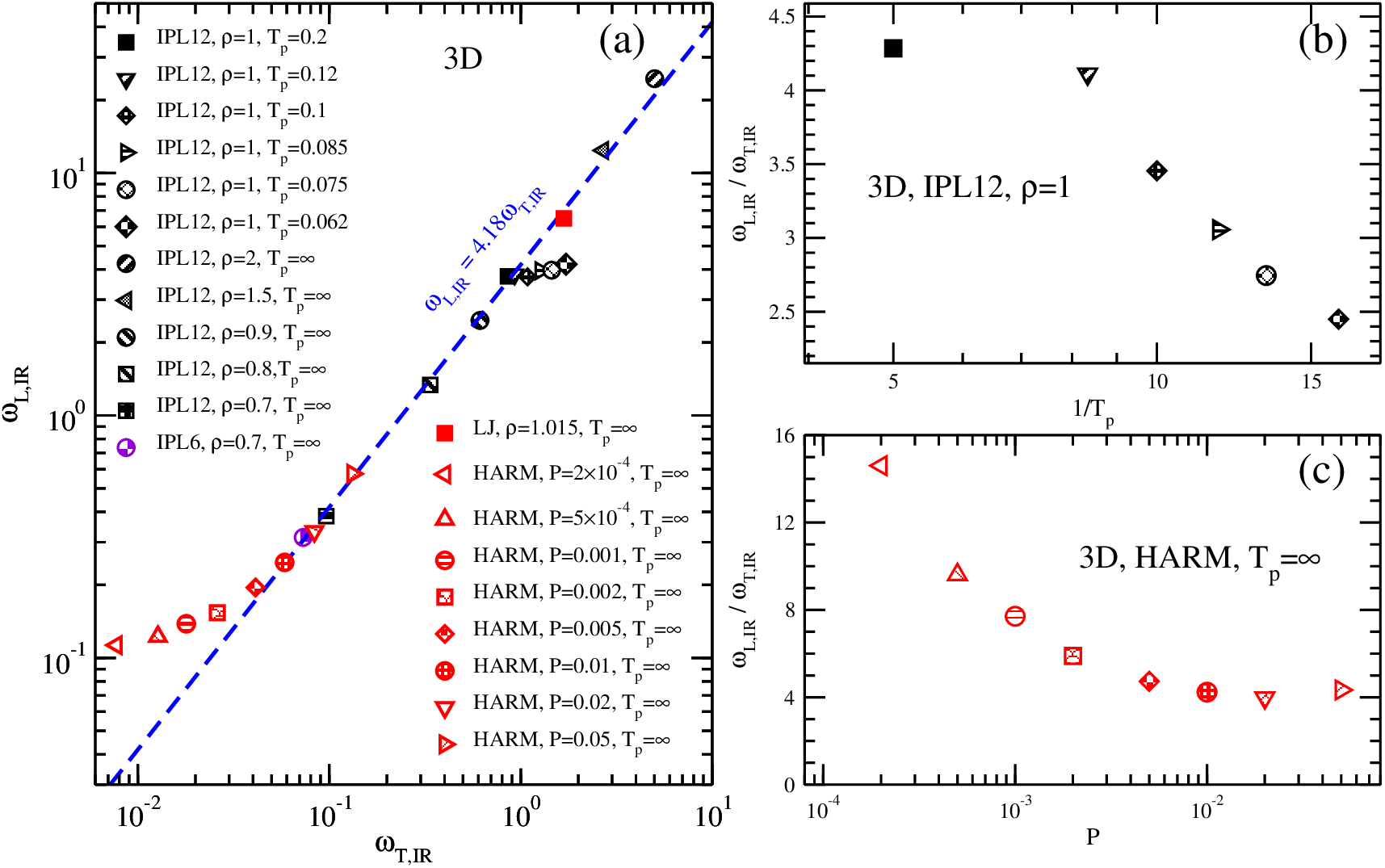}
\caption{\label{fig:3Dcompare} 
 (a) Correlation between the transverse IR frequency $\omega_{\rm T,IR}$ and the longitudinal IR frequency $\omega_{\rm L,IR}$ in 3D glasses. The data presented here were obtained from  model glass formers employing IPL,  HARM, and LJ  interaction potentials.  Data corresponding to LJ and HARM systems are extracted from   Refs.~\cite{Monaco-pnas-2009} and~\cite{Ikeda_pre2018}, respectively. The dashed line represents $\omega_{\rm L,IR}=4.18 \,\omega_{\rm T,IR}$.  
  (b) Parent temperature $T_{\rm p}$  dependence of the ratio $\omega_{\rm L,IR}/\omega_{\rm T,IR}$ in the (3D, IPL12, $\rho=1$) glasses.  Note that the glass stability increases with decreasing $T_{\rm p}$.    (c) Pressure $P$  dependence of the ratio $\omega_{\rm L,IR}/\omega_{\rm T,IR}$ in the (3D, HARM, $T_{\rm p}=\infty$) glasses. The jamming transition is at $P=0$, and systems with $P$ closer to this transition point tend to exhibit increased mechanical instability.   }
\end{figure*}

The sound attenuation coefficient $\Gamma_{\rm \lambda}(\omega)$ was extracted from the temporal decay of excited sound waves  within the harmonic approximation~\cite{Lemaitre-NM,Wang2019SMattenuation}. Here, $\lambda=\rm T$ and $\lambda=\rm L$ stand for transverse and longitudinal sound attenuation, respectively.  
First, the initial velocity of particle $i$ was prescribed as
\[
\mathbf{\dot{u}}_{i}(t=0)=\mathbf{b}_{\rm \lambda}\sin(\mathbf{k}\cdot\mathbf{r}_{i}) .
\]
Here, the unit vector $\mathbf{b}_{\rm \lambda}$ was chosen such that $\mathbf{b}_{\rm T}\cdot\mathbf{k}=0$ for transverse polarization and $\mathbf{b}_{\rm L}\mathbin{/\!/}\mathbf{k}$ for longitudinal polarization, where $\mathbf{k}$ is the wavevector, and $\mathbf{r}_{i}$ is the position of particle $i$ at $t=0$ in the $T=0$ glass. Second, molecular dynamics simulations were performed by integrating the  equations of motion, which read
\[
      \mathbf{ \ddot{u}}_{i}(t) =  \mathbf{ \dot{u} }_{i}(t=0) \delta(t)   -  \sum\limits^N \limits_{j=1}H_{ij} \mathbf{ u}_{j}(t),  
\]
where $H_{ij}$ is the Hessian matrix of $T=0$ glasses, and   $\mathbf{ u }_{i}(t)$  represents the displacement of particle $i$  at  $t$ relative to its initial position $\mathbf{r}_{i}$. We  calculated the velocity auto-correlation function,  
\[
 C_{\rm \lambda} (k,t) =\Bigg\langle
\frac{       \sum\limits^N \limits_{i=1}                 \mathbf   {\dot{ u }} _{i}(0)  \cdot              \mathbf   {\dot{ u }} _{i}(t)  }
{    \sum\limits^N \limits_{i=1}                 \mathbf   {\dot{ u }} _{i}(0)  \cdot              \mathbf   {\dot{ u }} _{i}(0)}\Bigg\rangle. 
\]
The  sound attenuation coefficient $\Gamma_{\rm \lambda}$ and  corresponding frequency $\omega$   could be  determined by  fitting the  $C_{\rm \lambda}(k,t)$ data to  
\begin{equation}
 C_{\rm \lambda} (k,t) =\exp(-\Gamma_{\rm \lambda} t/2)\cos(\omega t).
\label{eq1}
\end{equation}  

Figure~\ref{fig:envelope}  illustrates our procedure  to extract $\Gamma_{\rm T}$ and corresponding $\omega$ for transverse excitations.  Figure~\ref{fig:envelope}(a) presents the time evolution of $C_{\rm T}(k,t)$  at a small wavevector $k$. In principle, both $\Gamma_{\rm T}$ and $\omega$ can be  obtained by simply fitting the $C_{\rm T}(k,t)$ data to Eq.~\ref{eq1}.   In practice, this approach yields reliable estimates for $\omega$, but  performs poorly for $\Gamma_{\rm T}$ in finite-size systems, particularly at small $k$, because of the well-documented finite-size effects~\cite{Wang2019SMattenuation,Lemaitre-NM,lerner2019JCP}. These effects are more clearly manifested in the temporal evolution of the envelope of $C_{\rm T}(k,t)$, denoted as $E(k,t)$, as shown  in Fig.~\ref{fig:envelope}(b).   An exponential decay of $E(k,t)$ could be observed  over short to intermediate timescales, whereas deviations from the exponential behavior emerge at longer times~\cite{Wang2019SMattenuation,Lemaitre-NM,Fucpb}.  Nevertheless, recent simulation studies~\cite{Wang2019SMattenuation,Wang2020softmater,Fupre,Fucpb} have demonstrated that restricting the fit to the time window where the exponential decay applies is sufficient to determine  the sound attenuation coefficient.   Therefore, we employ the restricted envelope method to determine  the coefficient.  Unless stated otherwise, all reported $\Gamma_{\rm \lambda}(\omega)$ data were obtained from measurements performed on multiple system sizes. We have confirmed that the resulting $\Gamma_{\rm \lambda}(\omega)$  exhibits no apparent finite-size effects.

Figure~\ref{fig:OmegaIR}  illustrates how we determine the transverse IR frequency $\omega_{\rm T,IR}$ and the longitudinal IR frequency $\omega_{\rm L,IR}$. Specifically,   $\omega_{\rm T,IR}$ is determined such that $\pi\Gamma_{\rm T}(\omega_{\rm T,IR})=\omega_{\rm T,IR}$; analogously,  $\omega_{\rm L,IR}$ is defined by $\pi\Gamma_{\rm L}(\omega_{\rm L,IR})=\omega_{\rm L,IR}$.

\section{Results}

\subsection{Correlation between  glass stability and   longitudinal-to-transverse IR frequency ratio}

\begin{figure*}[t]
\includegraphics[width=0.9\textwidth]{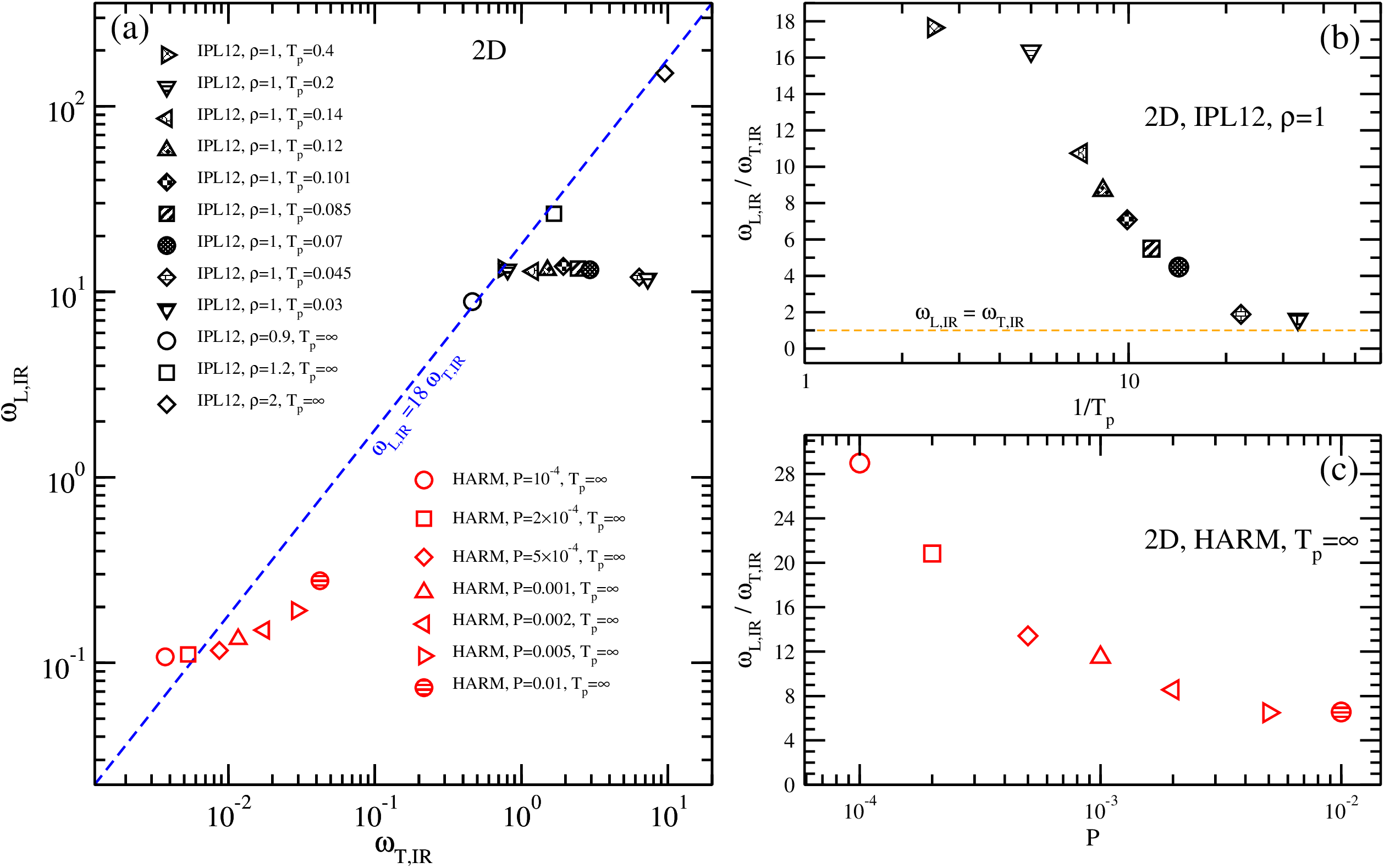}
\caption{\label{fig:2Dcompare} 
 (a) Correlation between  the transverse IR frequency $\omega_{\rm T,IR}$ and the longitudinal IR frequency $\omega_{\rm L,IR}$ in 2D glasses.  Data corresponding to HARM systems are extracted from Ref.~\cite{Ikeda_pre2018}. 
  (b) Parent temperature $T_{\rm p}$  dependence of the ratio $\omega_{\rm L,IR}/\omega_{\rm T,IR}$ in the (2D, IPL12, $\rho=1$) glasses.   (c) Pressure $P$  dependence of the ratio $\omega_{\rm L,IR}/\omega_{\rm T,IR}$ in the (2D, HARM, $T_{\rm p}=\infty$) glasses. 
}
\end{figure*}

Recent simulation studies  have  reported that $\omega_{\rm L,IR}$ is generally much greater  than $\omega_{\rm T,IR}$ across different  simple model glass formers~\cite{Ikeda_pre2018, bp_shintani_NM2008,  Monaco-pnas-2009,BPorigin_yuanchao-NP,xipeng-PRL}.  It is noteworthy that the glass samples investigated in these studies are obtained via rapid quenching of high-temperature liquids, and are therefore poorly annealed.  A natural question is whether $\omega_{\rm L,IR}$ necessarily remains substantially larger than $\omega_{\rm T,IR}$, or, alternatively, whether the ratio $\omega_{\rm T,IR}/\omega_{\rm L,IR}$ can vary under suitably chosen conditions.  To investigate this, we plot $\omega_{\rm L,IR}$ as a function of $\omega_{\rm T,IR}$ for 3D glasses in Fig.~\ref{fig:3Dcompare}(a) and for 2D glasses in Fig.~\ref{fig:2Dcompare}(a). The 3D IPL glasses considered here  differ in the  values of the exponent $n$ (with larger $n$ corresponding to a steeper repulsive interaction), the number density $\rho$, and the parent temperature $T_{\rm p}$. The data for (3D, LJ, $T_{\rm p}=\infty$) and (3D, HARM, $T_{\rm p}=\infty$) systems are extracted from the literature~\cite{Ikeda_pre2018,Monaco-pnas-2009}.

First,  as shown in Fig.~\ref{fig:3Dcompare}(a), for all poorly annealed 3D  glasses—except HARM glasses at low pressures—we observe that $\omega_{\rm L,IR} \approx 4.18\,\omega_{\rm T,IR}$, i.e.,  $\omega_{\rm L,IR} /\omega_{\rm T,IR} \approx 4.18\,\omega_{\rm T,IR}$,  irrespective of  interaction potential models.  A second regime where this linear scaling fails is observed in the (IPL12, $\rho=1$) glasses prepared with enhanced stability (parent temperature $T_{\rm p}<0.2$). In the following, we will discuss separately these two regimes where the relation $\omega_{\rm L,IR} \approx 4.18\,\omega_{\rm T,IR}$ is violated. Our strategy is to control all other variables while varying a single parameter at a time, thereby identifying the control parameter governing the ratio $\omega_{\rm L,IR}/\omega_{\rm T,IR}$. Notably, for poorly annealed 2D glasses, one  could also observed an approximately linear relationship between $\omega_{\rm L,IR}$ and $\omega_{\rm T,IR}$ over specific parameter regimes, see Fig.~\ref{fig:2Dcompare}(a).  However, unlike  the 3D case, the slope of the linear scaling in 2D case appears to depend  on the  interaction potential models.

Figure~\ref{fig:3Dcompare}(b) shows the dependence of $\omega_{\rm L,IR}/\omega_{\rm T,IR}$ on $1/T_{\rm p}$ for the (3D, IPL12, $\rho=1$) glasses. In this  system, glass stability increases as $T_{\rm p}$ decreases~\cite{wlj_dos_nc2019}, when $T_{\rm p}$ lies below $\approx 0.2$  (the onset temperature of slow dynamics). Note that  glasses prepared at $T_{\rm p}=0.2$ remain poorly annealed—comparable to glasses quenched from substantially higher $T_{\rm p}$—as evidenced by their nearly indistinguishable  properties  such as the sound attenuation~\cite{Wang2019SMattenuation} and  density of excess vibrational modes~\cite{wlj_dos_nc2019}. Across the full $T_{\rm p}$ range examined, we find that  $\omega_{\rm L,IR}/\omega_{\rm T,IR}$ decreases from approximately 4.3 to 2.4 as $1/T_{\rm p}$ increases. This trend indicates that enhanced stability reduces the difference  between $\omega_{\rm L,IR}$ and $\omega_{\rm T,IR}$, implying that annealing is an effective way to tune the ratio $\omega_{\rm L,IR}/\omega_{\rm T,IR}$.  Extrapolating this behavior would suggest that the two characteristic frequencies may ultimately coincide in   sufficiently stable  glasses; however, generating such stable samples is currently challenging in simulations~\cite{Ninarello_2026_JCP,Ninarello_2017_PRX}.  As an alternative route, the dependence of $\omega_{\rm L,IR}/\omega_{\rm T,IR}$ with the stability could be further assessed by applying the same analysis to 2D glasses prepared with different degrees of annealing~\cite{Wang2021prl,Berthier_2019_NC}.


We investigate the dependence of the ratio $\omega_{\rm L,IR}/\omega_{\rm T,IR}$ on $1/T_{\rm p}$ for the (2D, IPL12, $\rho=1$)  glasses, as shown in Fig.~\ref{fig:2Dcompare}(b). Consistent with the trend observed in 3D glasses, increasing glass stability leads to a reduction of the ratio $\omega_{\rm L,IR}/\omega_{\rm T,IR}$ in 2D glasses. Remarkably, $\omega_{\rm L,IR}/\omega_{\rm T,IR}$ decreases from approximately 17.6 in very poorly annealed  glasses ($T_{\rm p}=0.4$) to about 1.6 in the most stable samples studied here ($T_{\rm p}=0.03$). This pronounced evolution suggests that $\omega_{\rm L,IR}/\omega_{\rm T,IR}$ may approach unity in glasses with even higher stability. 

Two extrapolation scenarios are plausible in both 2D and 3D glasses: (i) $\omega_{\rm L,IR}/\omega_{\rm T,IR}=1$ is attained only in the limit of maximal stability, i.e., the so-called ideal glass~\cite{Berthier_2011_RMP,Fan-NM-ideal,Corwin-PRL-ideal}; or (ii) $\omega_{\rm L,IR}/\omega_{\rm T,IR}=1$ is reached already in glasses less stable than the ideal glass. In the latter case, an important open question is whether the ratio subsequently saturates at unity as the ideal-glass limit is approached, or whether it continues to decrease below 1 upon further stabilization. It is noteworthy that a ratio $\omega_{\rm L,IR}/\omega_{\rm T,IR} < 1$ was reported in one simulation study of a two-dimensional lattice model, where the disorder in lattice site positions was systematically tuned~\cite{Nie-Frontier}. 

Figures~\ref{fig:3Dcompare}(c) and~\ref{fig:2Dcompare}(c) show the pressure dependence of the ratio $\omega_{\rm L,IR}/\omega_{\rm T,IR}$ for the (3D, HARM, $T_{\rm p}=\infty$)  and (2D, HARM, $T_{\rm p}=\infty$) glasses, respectively.  The purely repulsive HARM potential has constituted  a prototypical model in the  study  of the jamming physics~\cite{Hecke—review, Andrea—review}. The jamming transition  occurring at $P=0$,  denotes a transition from a state without rigidity to a state  with rigidity, and the system at the  transition is thus marginally stable. For $P>0$, i.e., at finite distance from the jamming transition, the system moves away from marginality and becomes increasingly stable; accordingly, the mechanical stability increases  with increasing $P$.  We observe that $\omega_{\rm L,IR}/\omega_{\rm T,IR}$ decreases as $P$ increases, consistent with the trend observed  in 2D and 3D IPL glasses (see Figs.~\ref{fig:3Dcompare}(b) and~\ref{fig:2Dcompare}(b)) whose stability is tuned by varying $T_{\rm p}$. However,  when $P$ is greater than a crossover pressure $P_{\rm C} \approx 0.02 $,   the ratio $\omega_{\rm L,IR}/\omega_{\rm T,IR}$ seems to approach a plateau.  This behavior is obvious in  the (3D, HARM, $T_{\rm p}=\infty$) glasses;  for the (2D, HARM, $T_{\rm p}=\infty$) glasses, although there are not enough data in the high-pressure regime,   there seems to be a hint of saturation in the examined pressure regime. Apparently, the pressure-dependent behavior of the ratio    at pressures below and above  $P_{\rm C}$ exhibits distinct characteristics. 

Very recently,  one simulation study~\cite{Huang_NC_2026} suggests that in the HARM systems, there is a crossover pressure $P_{\rm Cref}$: the mechanical  behaviors  at pressures below $P_{\rm Cref}$ is mainly controlled by the bond connectivity  whose contribution increases with the distance to the jamming transition, while at pressures above  $P_{\rm C}$, the mechanical behaviors    are  mainly controlled by  internal stress, and the jamming physics plays minor role.  One consequence is that the mechanical stability cannot be inferred solely from the distance to the jamming transition in systems with  pressures above  $P_{\rm Cref}$, and hence the stability of  systems above  $P_{\rm Cref}$ may not increase with an increase in pressure. We observe that $P_{\rm C}$ and $P_{\rm C_{ref}}$ lie within the same regime, although whether they are identical requires further studies.  If  $\omega_{\rm L,IR}/\omega_{\rm T,IR}$ has a one-to-one correlation with the overall stability, the near invariance of $\omega_{\rm L,IR}/\omega_{\rm T,IR}$ for $P > P_{\rm C}$ would imply that the effective stability is approximately the same in this pressure regime. 

Additionally, in simulation studies of  glassy properties of generic  model glasses—particularly those employing purely repulsive interaction potentials—it is often asserted that the studied systems lie  far from the jamming transition to avoid the influence from  the jamming transition~\cite{lerner-JCP-review,lerner_prl2016,Wang-ropp,wang_3d_jcp2022}.     Our  results  indicate that upon entering the regime $P > P_{\rm C}$, the ratio of the longitudinal IR frequency to the transverse IR frequency is potentially uncorrelated with the jamming transition.   
From this perspective,  our results complement the proposition~\cite{Huang_NC_2026} that a crossover pressure exists that demarcates the jamming-physics-relevant regime from the non-jamming-physics-relevant regime. 


\begin{figure}[t]
\includegraphics[width=0.48\textwidth]{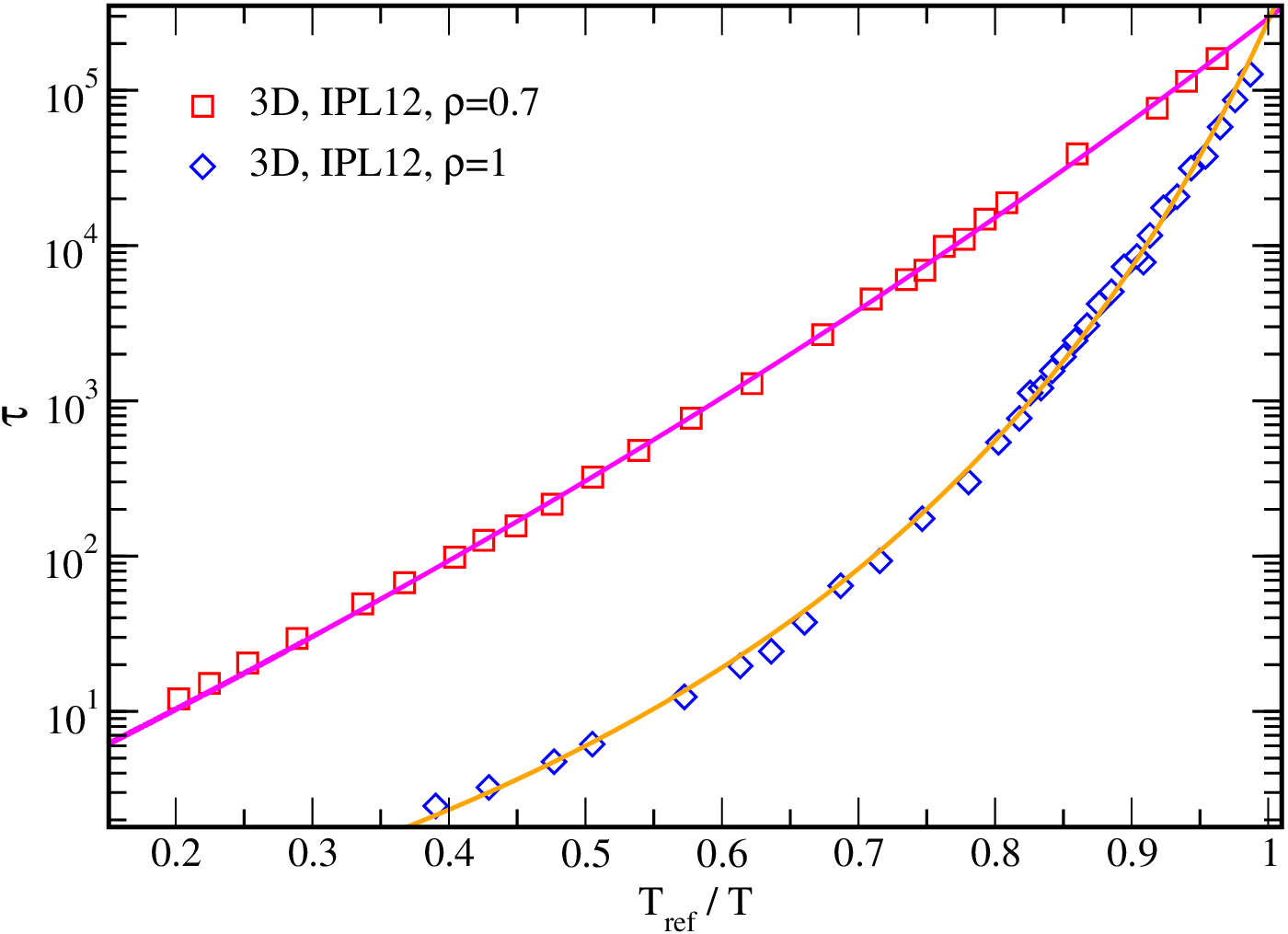}
\caption{\label{fig:Angell} Typical Angell plot of the structural relaxation time $\tau$ against the reciprocal  temperature $T_{\rm ref}/T$ in glass-forming liquids. Here,  $T_{\rm ref}$ is defined such that $\tau(T_{\rm ref})=2\times 10^{5}$.  The solid curves represent fitting to the Vogel-Fulcher-Tammann equation~\cite{Berthier_2011_RMP}. 
}
\end{figure}

\begin{figure*}[t]
\includegraphics[width=0.98\textwidth]{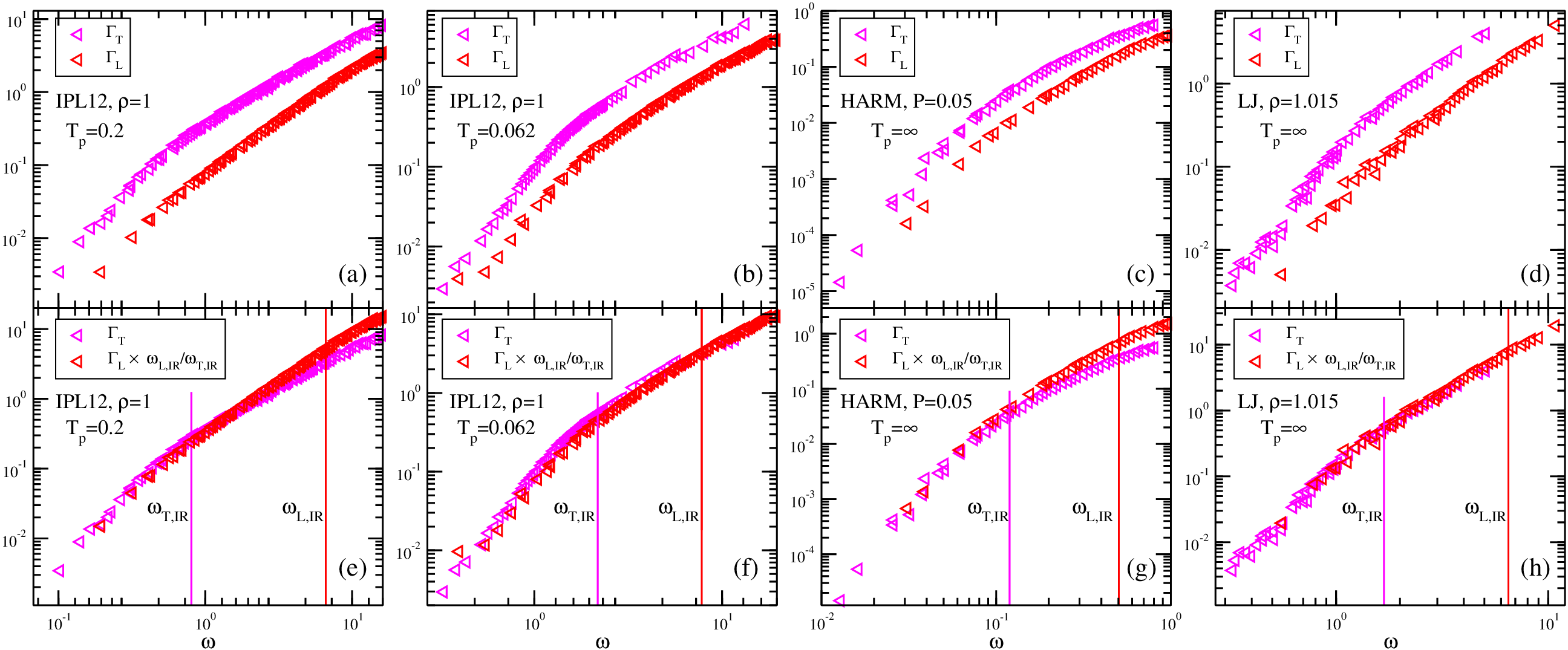}
\caption{\label{fig:3Dscaling} Demonstration of the low-frequency relationship, $\omega_{\rm T,IR} \,\Gamma_{\rm T}(\omega)=\omega_{\rm L,IR}\, \Gamma_{\rm L}(\omega)$, for four representative  3D glassy samples.    Panels (a)–(d) display $\Gamma_{\rm T}(\omega)$ and $\Gamma_{\rm L}(\omega)$. Panels (e)–(h) present  $\Gamma_{\rm T}(\omega)$ and   $(\omega_{\rm L,IR}/\omega_{\rm T,IR})\,\Gamma_{\rm L}(\omega)$.  Panels (a) and (e) correspond to  identical glasses as marked, and the same correspondence applies to pairs (b) and (f), (c) and (g), as well as (d) and (h).    The  values of $\omega_{\rm L,IR}/\omega_{\rm T,IR}$ in panels (e), (f),  (g),  and (h) are 4.36482, 2.43870, 4.21867 and 3.86519, respectively.  The system-specific IR frequencies $\omega_{\rm T,IR}$ and $\omega_{\rm L,IR}$ are indicated by magenta and red vertical lines, respectively,  in panels (e)–(h).  Data corresponding to (LJ, $\rho=1.015$, $T_{\rm p}=\infty$) and  (HARM, $P=0.05$, $T_{\rm p}=\infty$) systems are extracted from   Refs.~\cite{Monaco-pnas-2009} and~\cite{Ikeda_pre2018}, respectively.
}
\end{figure*}

\subsection{Correlation between liquid fragility  and   longitudinal-to-transverse IR frequency ratio}

Recent computer simulation studies indicate that, in silica glasses, $\omega_{\rm L,IR}=\omega_{\rm T,IR}$~\cite{silica-model-1,silica-model-2}. A  widely documented property with respect to silica glasses is that their parent glass-forming liquids are classified as “strong” in the sense of Angell plot~\cite{Staley-JCP-2015,Berthier_2011_RMP,Angell_1991_JNCS,Angell_1995_Science}. In the Angell plot, the  structural relaxation time $\tau$ is represented as a function of $T_{\rm ref}/T$, as illustrated in Fig.~\ref{fig:Angell}. Here, $T_{\rm ref}$ denotes a system-dependent reference temperature defined such that the corresponding $\tau$ reaches a prescribed large value.  In glass-forming liquids, $\tau$ generally increases sharply upon cooling toward the glass transition, even for comparatively small reductions in $T$. However, the magnitude of this temperature sensitivity varies substantially across different glass-forming liquids. For some systems, the temperature dependence of $\tau$ follows the Arrhenius law~\cite{Staley-JCP-2015,Berthier_2011_RMP}, $\tau \sim \exp(E/T)$, with a temperature-independent activation energy $E$. In such cases, the $\tau$ versus $T_{\rm ref}/T$ representation yields a straight line in the Angell plot, and the liquid is termed “strong”. By contrast, when $\tau(T)$ deviates from the Arrhenius behavior—the activation energy  increases upon decreasing $T$—the corresponding Angell-plot curve   exhibits typically upward curvature, and the liquid is termed “fragile”:   more fragile systems exhibit greater deviations from the Arrhenius behavior. 


 Given that silica glass-forming liquids are  strong~\cite{Staley-JCP-2015}  and  that $\omega_{\rm L,IR}=\omega_{\rm T,IR}$~\cite{silica-model-1,silica-model-2} in silica glasses, it is natural to ask whether there exists a systematic connection between liquid fragility and the ratio $\omega_{\rm L,IR}/\omega_{\rm T,IR}$ measured in the corresponding glass. First, we examine whether variations in the fragility of the parent glass-forming liquids influence $\omega_{\rm L,IR}/\omega_{\rm T,IR}$ in the resulting poorly annealed glasses. Note that there have been some studies~\cite{Novikov-Poisson-Nature,experiment-JCP-Ruta-IRgtBP,Wang_prl2014}  attempting to make direct connections between properties of glasses and corresponding glass-forming liquids.  We consider the (3D, IPL12, $\rho=0.7$, $T_{\rm p}=\infty$) and (3D, IPL12, $\rho=1$, $T_{\rm p}=\infty$) glasses, and analyze the dynamics of their corresponding glass-forming liquids.
Figure~\ref{fig:Angell} presents the Angell plots for the (3D, IPL12, $\rho=1$) and (3D, IPL12, $\rho=0.7$) glass-forming liquids. It can be observed that the curve corresponding to the (3D, IPL12, $\rho = 1$) system exhibits substantially greater curvature than that of the (3D, IPL12, $\rho = 0.7$) system,   indicating vastly different fragility. Despite this pronounced difference, the ratios $\omega_{\rm L,IR}/\omega_{\rm T,IR}$ for the (3D, IPL12, $\rho=0.7$, $T_{\rm p}=\infty$)  and (3D, IPL12, $\rho=1$, $T_{\rm p}=\infty$) glasses are approximately identical, see Fig.~\ref{fig:3Dcompare}(a). The above observations  suggest that changes in  fragility in glass-forming liquids are not directly reflected in changes of $\omega_{\rm L,IR}/\omega_{\rm T,IR}$ in corresponding glasses.

A complementary question is whether strong glass-forming liquids necessarily yield $\omega_{\rm L,IR}=\omega_{\rm T,IR}$ in the corresponding poorly annealed glasses. For the (3D, IPL12, $\rho=0.7$) liquids, the curve in the Angell plot is  approximately  straight, which is indicative of  quasi-strong glass-forming behavior.
Nevertheless, the associated glasses (3D, IPL12, $\rho=0.7$, $T_{\rm p}=\infty$) do not satisfy $\omega_{\rm L,IR}\approx \omega_{\rm T,IR}$.  This further indicates  that the fragility of the glass-forming liquid exhibits no clear or direct correlation with the ratio $\omega_{\rm L,IR}/\omega_{\rm T,IR}$ in the corresponding glasses.  However, in network silica glasses,  the corresponding glass-forming liquids are strong, and it is indeed  observed that  $\omega_{\rm L,IR}=\omega_{\rm T,IR}$.  This observation of the equality  is likely attributable to a high degree  stability conferred by the strong covalent bonding in silica, given that glass stability controls the ratio $\omega_{\rm L,IR}/\omega_{\rm T,IR}$ from discussions in Figs.~\ref{fig:3Dcompare} and ~\ref{fig:2Dcompare}. It is worth noting that further research is required to develop an effective parameter capable of quantitatively characterizing the overall stability of diverse glasses, including those constructed with different interaction potentials, to facilitate the comparison of glass stability under a unified standard. 

\subsection{Correlation  between    longitudinal-to-transverse IR frequency ratio and    low-frequency transverse-to-longitudinal sound attenuation coefficient ratio}

Although it seems that the literature more extensively addresses transverse sound attenuation than longitudinal  attenuation, several studies have nevertheless reported a quantitative relationship between the corresponding attenuation coefficients at low frequencies, namely $\Gamma_{\rm T}(\omega)/\Gamma_{\rm L}(\omega)=\beta_{\rm T/L,\Gamma}$, where $\beta_{\rm T/L,\Gamma}$ is a system-specific constant. To our knowledge, this proportionality was first identified numerically in 3D LJ glasses~\cite{Monaco-pnas-2009} (the same data are extracted to serve as LJ data adopted  in our study)  and was subsequently verified in 2D and 3D glasses spanning different degrees of stability~\cite{Wang2019SMattenuation,Wang2020softmater,Fupre}. 
In these works,  $\beta_{\rm T/L,\Gamma}$ was selected empirically  to collapse the low-frequency $\Gamma_{\rm T}(\omega)$ data onto the rescaled longitudinal data, $[\beta_{\rm T/L,\Gamma}\,\Gamma_{\rm L}(\omega)]$. However, it  remains unclear whether $\beta_{\rm T/L,\Gamma}$ can be determined in a  quantitatively predictive manner.

Here, we observe that, for our studied 3D  glasses, $\beta_{\rm T/L,\Gamma}\approx \omega_{\rm L,IR}/\omega_{\rm T,IR}$. This is demonstrated in Fig.~\ref{fig:3Dscaling} for representative systems that differ in  interaction potential or stability. In particular, the transverse attenuation $\Gamma_{\rm T}(\omega)$ overlaps with the rescaled longitudinal attenuation $[(\omega_{\rm L,IR}/\omega_{\rm T,IR})\,\Gamma_{\rm L}(\omega)]$ at low frequencies extending  up to at least   $\omega_{\rm T,IR}$. These results indicate that, in 3D glasses, $\beta_{\rm T/L,\Gamma}$ and $\omega_{\rm L,IR}/\omega_{\rm T,IR}$ are governed by the same control parameters.  In contrast, for 2D glasses we do not find a comparable coincidence between $\beta_{\rm T/L,\Gamma}$ and $\omega_{\rm L,IR}/\omega_{\rm T,IR}$, although, like $\omega_{\rm L,IR}/\omega_{\rm T,IR}$,   $\beta_{\rm T/L,\Gamma}$ also  decreases  with increasing glass stability.    Instead,  we observe that the relationship in 2D glasses evolves systematically with stability, changing from $\beta_{\rm T/L,\Gamma} \ll \omega_{\rm L,IR}/\omega_{\rm T,IR}$ in very poorly annealed glasses to $\beta_{\rm T/L,\Gamma} > \omega_{\rm L,IR}/\omega_{\rm T,IR}$ in very stable glasses.

\section{Conclusions}

In summary, we investigate the quantitative relationship between the longitudinal and transverse IR frequencies in 2D and 3D model glasses. We find that glass stability constitutes a key control parameter for the longitudinal-to-transverse IR frequency ratio in both 2D and 3D systems: this ratio decreases systematically with increasing stability and reaches approximately 1.6 in the most stable 2D glasses under study. Although additional work is needed to establish whether a lower bound exists for this stability-dependent evolution, our results suggest that the transverse and longitudinal IR frequencies may approach coincidence. If, as commonly observed, the boson peak frequency coincides with the transverse IR frequency~\cite{bp_shintani_NM2008}, irrespective of glass stability, then it may also coincide with the longitudinal IR frequency in sufficiently stable glasses. In that scenario, longitudinal modes would contribute materially to elucidating the  origin of the boson peak. This possibility contrasts with the prevailing practice of neglecting longitudinal modes, which is largely based on the empirical observation that the longitudinal IR frequency is typically much higher than its transverse counterpart.

Second, we do not identify a clear correlation between the fragility of glass-forming liquids and the longitudinal-to-transverse IR frequency ratio in the corresponding glasses. This observation suggests that the near equality of longitudinal and transverse IR frequencies in silica glass is unlikely to be attributable to silica glass-forming liquids being “strong” in the Angell classification.

Final, we find that, in 3D glasses, the ratio of the longitudinal to transverse IR frequency is approximately equal to the ratio of the transverse to longitudinal sound attenuation coefficient at a fixed low frequency, at least for frequencies not exceeding the transverse IR frequency, i.e., $\omega_{\rm T,IR}\,\Gamma_{\rm T}(\omega)=\omega_{\rm L,IR}\,\Gamma_{\rm L}(\omega)$. Note that  this relation does not appear to hold in 2D glasses. If this relation were to remain valid in more complex glassy systems, it would offer a practical strategy for estimating the low-frequency  \(\Gamma_{\rm T}(\omega)\) in 3D experimental glasses, which is not directly accessible in   experimental measurements~\cite{Wang-PRL-quartic-2025, Nakayama-ROPP-2002,BPorigin-Rufﬂe-PRL-TwoModels,experiment-JCP-Ruta-IRgtBP,experiment-PRB-Ramos-IRgtBP,experiment-PRL-Baldi-IReqBP}. Specifically,  \(\Gamma_{\rm L}(\omega)\), the corresponding \(\omega_{\rm L,IR}\), and the boson peak frequency are, in general, experimentally accessible. By invoking the widely reported empirical coincidence between the boson peak frequency and \(\omega_{\rm T,IR}\)~\cite{bp_shintani_NM2008}, one may subsequently infer \(\Gamma_{\rm T}(\omega)\) directly from the relation given above. Nevertheless, it deserves further work  to elucidate the underlying mechanism of this relation and verify its generality.

\section*{ACKNOWLEDGMENTS}

We wish to thank Andrea Ninarello for generously providing equilibrated configurations at  low parent temperatures. We acknowledge  the support from   National Natural Science Foundation of China (Grant Nos. 12522503 and 12374202) and Anhui Projects (Grant Nos. 2022AH020009, S020218016,  and Z010118169).  We  also acknowledge Hefei Advanced Computing Center, Beijing Super Cloud Computing Center, and the High-Performance Computing Platform of Anhui University for providing computing resources.



\end{document}